\documentclass[twoside,twocolumn]{article}

\usepackage{blindtext} 

\usepackage[sc]{mathpazo} 
\usepackage[T1]{fontenc} 
\usepackage{microtype} 

\usepackage[english]{babel} 

\usepackage[hmarginratio=1:1,top=32mm,columnsep=20pt]{geometry} 
\usepackage[footnotesize]{caption} 
\usepackage{enumitem} 
\setlist[itemize]{noitemsep} 

\usepackage{abstract} 

\usepackage{titlesec} 
\titleformat*{\section}{\large\scshape\bfseries}
\titleformat*{\subsection}{\normalsize\bfseries} 
\titleformat*{\subsubsection}{\normalsize\it} 

\usepackage{titling} 

\usepackage{cite}
\usepackage{epstopdf}
\usepackage{graphicx}
\usepackage{float}
\usepackage{subfigure}
\usepackage{rotating}
\usepackage{pdflscape}
\usepackage{multirow}
\newcommand{\tabincell}[2]{\begin{tabular}{@{}#1@{}}#2\end{tabular}}
\usepackage{url}
\usepackage[colorlinks,
            linkcolor=blue,
            anchorcolor=blue,
            citecolor=blue]{hyperref}
\usepackage{amsmath}
\usepackage{algorithm}
\usepackage{algorithm}
\usepackage[noend]{algpseudocode}

\pretitle{\begin{center}\Large} 
\posttitle{\end{center}} 
\title{Identification of Autism spectrum disorder based on a novel feature selection method and Variational Autoencoder} 
\usepackage{authblk}
\author[a]{Fangyu Zhang}
\author[b]{Yanjie Wei}
\author[d]{Jin Liu}
\author[b]{Yanlin Wang}
\author[b]{Wenhui Xi}
\author[b,c]{Yi Pan \thanks{Corresponding author: Yi Pan (yi.pan@siat.ac.cn)}}
\affil[a]{College of Engineering, Southern University of Science and Technology, Shenzhen, 518055, China}
\affil[b]{Centre for High Performance Computing, Shenzhen Institutes of Advanced Technology, Chinese Academy of Sciences, Shenzhen, 518055, China}
\affil[c]{College of Computer Science and Control Engineering, Shenzhen Institutes of Advanced Technology, Chinese Academy of
Sciences, Shenzhen, 518055, China}
\affil[d]{School of Information Science and Engineering, Central South University, Changsha, 410083, China}

\date{}
\renewcommand{\maketitlehookd}{%
\begin{abstract}
The development of noninvasive brain imaging such as resting-state functional magnetic resonance imaging (rs-fMRI) and its combination with AI algorithm provides a promising solution for the early diagnosis of Autism spectrum disorder (ASD). However, the performance of the current ASD classification based on rs-fMRI still needs to be improved. This paper introduces a classification framework to aid ASD diagnosis based on rs-fMRI. In the framework, we proposed a novel filter feature selection method based on the difference between step distribution curves (DSDC) to select remarkable functional connectivities (FCs) and utilized a multilayer perceptron (MLP) which was pretrained by a simplified Variational Autoencoder (VAE) for classification. We also designed a pipeline consisting of a normalization procedure and a modified hyperbolic tangent (tanh) activation function to replace the original tanh function, further improving the model accuracy. Our model was evaluated by 10 times 10-fold cross-validation and achieved an average accuracy of 78.12\%, outperforming the state-of-the-art methods reported on the same dataset. Given the importance of sensitivity and specificity in disease diagnosis, two constraints were designed in our model which can improve the model's sensitivity and specificity by up to 9.32\% and 10.21\%, respectively. The added constraints allow our model to handle different application scenarios and can be used broadly. \\\\
\it{Keywords:} ASD; fMRI; Filter feature selection; VAE; ABIDE; Classification
\end{abstract}
}

\begin{document}

\maketitle

%
%
%

\section{Introduction}
\label{s1}
Autism spectrum disorder (ASD) is a common complex neurodevelopmental disorder that occurs in early childhood and has received much attention in recent years because of its high incidence and difficulty in curing. Although the manifestation of individuals with ASD varies greatly with age and ability \cite{1}, the disorder is characterized by core features in two areas—social communication and restricted, repetitive sensory-motor behaviors \cite{2}. Some ASD patients can go undetected during childhood and be observed to have psychiatric comorbidity during adolescence \cite{aggarwal2015misdiagnosis}, which presents challenges for traditional symptom-based diagnostic methods in ASD early diagnosis. Studies have shown that multiple biological factors can lead to the same ASD-related behavioral phenotype \cite{32}. However, traditional symptom-based ASD detection methods are unable to give a reliable diagnosis from a pathogenic perspective. To bridge this gap, non-invasive brain imaging techniques such as resting-state functional Magnetic Resonance Imaging (rs-fMRI) have been used to reveal valuable information about brain network organization \cite{PLIS20111156} and contribute to a better understanding of the neural circuitry underlying ASD and its associated symptoms. By measuring changes in Blood Oxygen level-dependent (BOLD) signals, rs-fMRI can detect the functional connectivity patterns between the brain regions of interest (ROIs), and several studies have found abnormal functional connectivities in ASD subjects \cite{hull2017resting, supekar2013brain, just2012autism}. With the development of artificial intelligence technology, deep learning techniques have made it possible to process and analyze large amounts of fMRI data to discover patterns amongst the complex functional connectivities that are not apparent to the human eye and have achieved good results for ASD prediction, making fMRI-based deep learning models a promising auxiliary tool for ASD early screening. 

However, small sample sizes and high feature dimensionality increase the challenge of developing a robust and well-performed machine learning model. One subject's feature vector composed of functional connectivities (FCs) extracted from rs-fMRI usually has tens of thousands of dimensions which makes the model training a time-consuming process and contains a lot of noise resulting from the recording image process \cite{KHODATARS2021104949} and redundancy information which will affect the model performance. Feature selection is a dimensionality reduction technique that identifies the key features of a given problem \cite{REMESEIRO2019103375}. The main goal of feature selection is to construct a subset of features as small as possible, which represent the vital features of the entire input data \cite{Zebari_Abdulazeez_Zeebaree_Zebari_Saeed_2020}. As an effective method to reduce training time and improve model performance, feature selection has been widely used in previous studies \cite{10.3389/fnins.2017.00460, WANG201999, 13, li2020selection, latkowski2015data, washington2019feature}. In the work of Guo et al. \cite{10.3389/fnins.2017.00460}, a feature selection method based on multiple sparse autoencoders achieved a 9.09\% model accuracy improvement based on 55 ASD and 55 Healthy Controls (HC) subjects from UM site of Autism Brain Imaging Data Exchange (ABIDE) I dataset \cite{abide}. Wang et al. \cite{WANG201999} searched for informative features by SVM-RFE and obtained 90.60\% accuracy by SVM classifier on a dataset consisting of 255 ASD and 276 HC subjects. In \cite{13}, a graph-based feature selection method was proposed to select remarkable FCs. Based on the refined features, a deep belief network (DBN) was trained and achieved a higher classification accuracy (76.4\%) than previous studies on the entire ABIDE I dataset. 

In previous studies, machine learning algorithms such as support vector machine (SVM), decision tree, and Gaussian naive Bayes have been applied to ASD recognition \cite{9, 10}, most of which belong to supervised learning methods. With the development of deep learning technology, pretraining classifiers using unsupervised deep learning methods has been proved to be helpful to improve the performance of classifiers. For instance, Heinsfeld et al. \cite{11} utilized a denoising autoencoder to extract underlying representations of the input feature vectors and then trained fully connected layers for classifying ASD from HC based on 1035 subjects from the ABIDE I dataset. Their model achieved better classification accuracy (70\%) than SVM and random forest (RF). Kong et al. \cite{31} pretrained their model by a sparse autoencoder based on 182 subjects from NYU site of ABIDE I and achieved an accuracy of 90.39\% and the area under the receiver operating characteristic curve (AUC) of 0.9738 for ASD/HC classification. Although denoising autoencoders and sparse autoencoders are widely used in previous studies and got relatively good results, the setting of the noise ratios and sparsity parameters are subjective to some extent. Some other researchers such as Sherkatghanad et al. \cite{30} and Shrivastava et al. \cite{28} performed ASD classification by using CNN to process the functional connectivity matrix. These CNN-based methods can extract local characteristics of images, however, the functional connectivity matrix is not an image in Euclidean space and doesn't have specific local characteristics. Other state-of-the-art methods that have been used in ASD classification include DBN \cite{13}, CapsNet \cite{29}, ASD-Diagnet \cite{12}, etc., all of these methods have achieved over 70\% prediction accuracy. Given that the doctor's diagnosis results can be affected by subjective factors such as different clinical experiences and fatigue, as well as objective factors such as the patient's insignificant symptoms, false negative and false positive cases have always been difficult to avoid, while most previous studies primarily concentrated on developing high-accuracy models without taking measures to improve sensitivity or specificity, which are critical for reducing the false-negative rate and false-positive rate, respectively. 

The purpose of this paper is to provide an ASD/HC classification framework based on rs-fMRI to improve ASD classification performance on heterogeneous datasets and to flexibly improve model sensitivity or specificity according to actual needs. In the framework, we proposed a novel feature selection method based on the difference between step distribution curves (DSDC) that not only contributed to higher classification accuracy but significantly speed up the training process. To get a more accurate and reliable ASD/HC classification model, we simplified the architecture of Variational Autoencoder (VAE) to pretrain the classifier and designed a pipeline consisting of a normalization and a modified hyperbolic tangent (tanh) activation function to replace the original tanh activation function, and adopted the threshold moving approach which is described in Section \ref{s2.4.3} to alleviate the impact of class imbalance of the dataset. In addition, we designed two constraints, by using which in the training process, model sensitivity or specificity can be effectively improved. The proposed method can potentially be used for ASD early screening and provide a valuable reference for doctors' decision-making.

Our main contributions are summarized as follows:

1. We proposed an ASD/HC classification framework including a novel feature selection method (the DSDC-based feature selection method), a simplified VAE pretraining method, and an MLP classifier. The accuracy of our classifier outperforms the state-of-the-art results reported on the same dataset with an outstanding training speed.

2. We designed two constraints that can be used during the model training process to effectively improve model sensitivity and specificity, respectively. 

This paper is structured as follows: Section \ref{s2} introduces the dataset we used and rs-fMRI data preprocessing process (\ref{s2.1}-\ref{s2.2}), feature selection method (\ref{s2.3}) and details of our model (\ref{s2.4}-\ref{s2.5}). Section \ref{s3} discusses the experimental results and limitations. Finally, the conclusion and future work are presented in section \ref{s4}.


\section{Materials and methods}
\label{s2}
\subsection{Participants} 
\label{s2.1}
ABIDE I dataset is one of the most commonly used public datasets taken from 17 international sites (\url{http://preprocessed-connectomes-project.org/abide/}). In order to train a robust model with stronger generalization ability for the data from different sites, our study was carried out using all valid rs-fMRI data from ABIDE I including 505 ASD and 530 HC samples, the largest rs-fMRI subset of ABIDE I that has ever been used, the phenotype of which is summarized in Table \ref{Table 1}.

\begin{table}[h!]
\renewcommand\arraystretch{1.5}  
\scriptsize
\caption{Demographic description of participants for ABIDE I}
{\tabcolsep0.07in  
  \begin{tabular}{lcccccc}
	\hline
	\textbf{Site} & \textbf{ASD} & \textbf{HC} & \textbf{Male} & \textbf{Female} & \textbf{Subtotal} & \textbf{Average age} \\
	\hline
	CALTECH &	19&	18&	29&	8&	37&	27\\
	CMU &	     14&	13&	21&	6&	27&	26\\
	KKI &	     20&	28&	36&	12&	48&	10\\
	LEUVEN &	29&	34&	55&	8&	63&	18\\
	MaxMun &	24&	28&	48&	4&	52&	25\\
	NYU &	     75&	100&	139&	36&	175&	15\\
	OHSH &	12&	14&	26&	0&	26&	10\\
	OLIN &	19&	15&	29&	5&	34&	16\\
	PITT &	29&	27&	48&	8&	56&	18\\
	SBL &	      15&	15&	30&	0&	30&	34\\
	SDSU &	14&	22&	29&	7&	36&	14\\
	Stanford &	19&	20&	31&	8&	39&	9\\
	Trinity &	22&	25&	47&	0&	47&	16\\
	UCLA &	54&	44&	86&	12&	98&	13\\
	UM &	     66&	74&	113&	27&	140&	14\\
	USM &	     46&	25&	71&	0&	71&	22\\
	Yale &	28&	28&	40&	16&	56&	12\\
	Total:&	505&	530&	878&	157&	1035	\\
	\hline
  \end{tabular}
}
\label{Table 1}
\end{table}

\begin{figure*}[h!]
\includegraphics[scale =0.32]{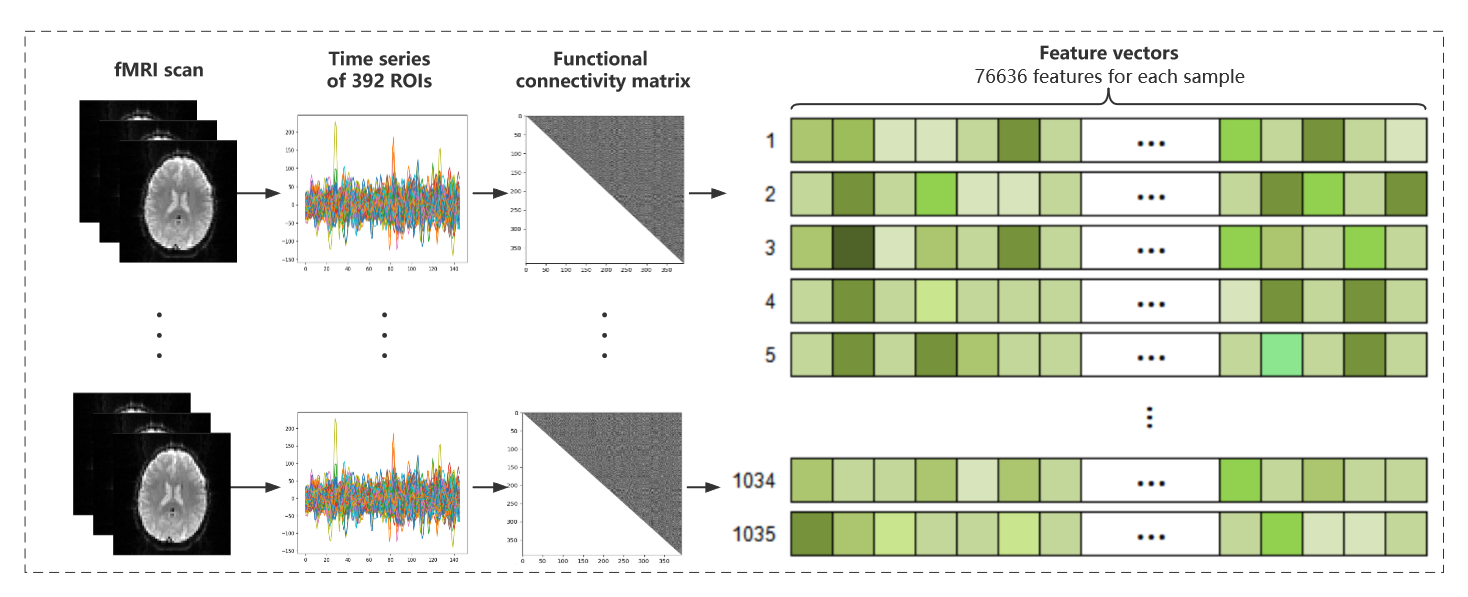}
\caption{The process of generating subjects' feature vectors from fMRI images}
\label{fig1}
\end{figure*}

Four different preprocessed datasets have been provided by ABIDE I according to four pipelines (CPAC \cite{18}, CCS, DPARSF, and NIAK). The CPAC pipeline was considered in our work, partly because previous studies such as Zhang et al. \cite{19} have compared the four different pipelines and found that data preprocessed with CPAC pipeline can achieve better classification results, and partly because it allows our model to be compared and evaluated against most of the other methods that have chosen CPAC pipeline as well.
In addition, ABIDE I dataset provides data preprocessed by seven brain atlases among which Craddock 200 (CC200) \cite{20} and Craddock 400 (CC400) have been proved by previous studies better than other atlases such as Automated Anatomical Labeling(AAL), Dosenbach160, etc \cite{22, 12}. CC400 atlas was adopted in our work mainly because CC400 has a more detailed division of ROIs than CC200. Another important reason is that the DSDC-based feature selection greatly reduced the dimension of the original feature vector, which enables us to complete the model training in a short time even based on a complex brain atlas like CC400 (392 ROIs, 76636 features for each subject).

\subsection{Functional Connectivity Measures and Subject’s Feature Vector}
\label{s2.2}
CC400 is a brain atlas with 392 ROIs from which 392 time series were extracted. The Pearson correlation coefficient (PCC) was calculated between each pair of time series by Formula (\ref{eq:pcc}) to measure the coactivation level between each pair of ROIs and form a PCC matrix. For each subject, we took the upper triangle of the PCC matrix and removed the main diagonal elements, after which the remaining triangle was flattened into a one-dimensional feature vector including $392 \times (392 - 1) / 2 = 76636$ features. The entire process of generating subjects' feature vectors is shown in Figure \ref{fig1}.
\begin{equation}
\begin{aligned}[b]
PCC_{x,y}=\frac{\sum_{i=1}^{N}(x_i-\Bar{x})(y_i-\Bar{y})}{\sqrt{\sum_{i=1}^{N}(x_i-\Bar{x})^2}\sqrt{\sum_{i=1}^{N}(y_i-\Bar{y})^2}}
\label{eq:pcc}
\end{aligned}
\end{equation}
Where $PCC_{x,y}$ is the PCC between time series x and y; N is the length of time series; $\Bar{x}$ and $\Bar{y}$ are the mean value of time series x and y. 

\subsection{DSDC-based feature selection}
\label{s2.3}
\begin{figure*}[h!]
\centering
\includegraphics[scale=0.73]{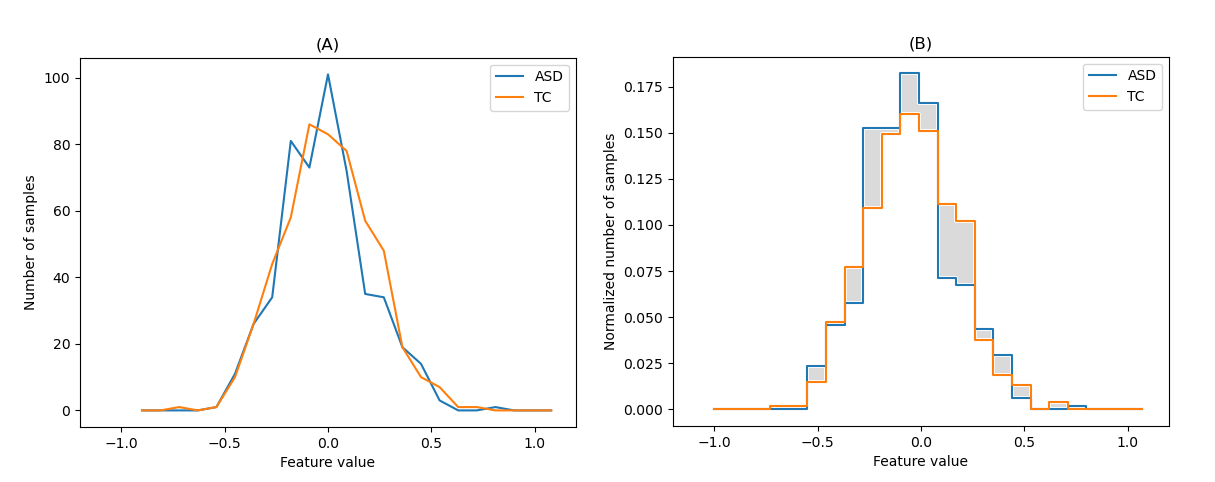} 
\caption{The original distribution curves and the corresponding step distribution curves for the functional connectivities between Right Middle Frontal Gyrus and Right Superior Frontal Gyrus. (A) Original distribution curves; (B) Step distribution curves}
\label{fig2}
\end{figure*}

A subject can be represented by a feature vector as explained in section \ref{s2.2}, while most features have complex distributions across subjects. In order to reduce the complexity of the distribution curves, the value range of features was evenly divided into 20 subintervals. Within each subinterval, for each class, the number of samples was divided by the total sample size of the class and got a normalized value. By using all the normalized values, step distribution curves (see Figure \ref{fig2} (B)) were created to approximate the original feature distribution curves (see Figure \ref{fig2} (A)). We defined the DSDC score in equation (2) to measure the distribution difference of positive and negative samples.

\begin{equation}
\begin{aligned}[b]
DSDC\_score=\sum_{i=b_0+\delta}^{b_1}|n_i^+/N^+ - n_i^-/N^-|
\label{eq:dsdc}
\end{aligned}
\end{equation}
Where $b_0$ and $b_1$ represent the lower bound and upper bound of feature values; $\delta$ is  the span of subinterval; $n_i^+$ and $n_i^-$
are the numbers of positive and negative samples whose feature values are in $[i-\delta, i)$; $N^+$ and $N^-$ are the positive and negative sample sizes. The larger the DSDC score of a feature, the more discriminative the feature is. 

\begin{figure}[h!]
\includegraphics[scale=0.5]{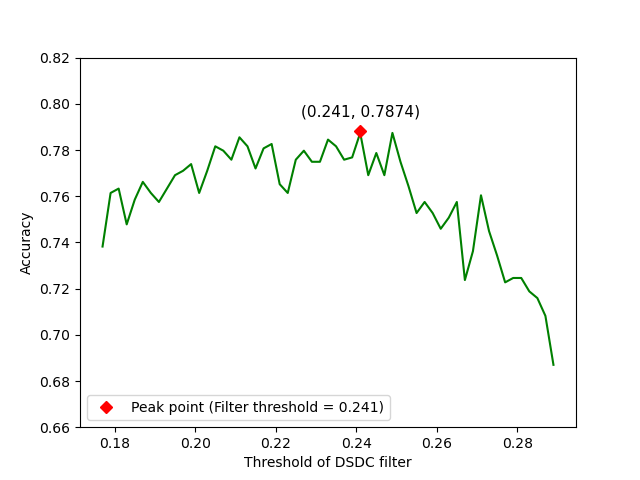} 
\caption{Selecting the filter threshold corresponding to the highest average accuracy of the 10-fold cross-validation}
\label{fig3}
\end{figure}

In our experiment, a feature was considered remarkable if its DSDC score is bigger than a preset filter threshold (0.241). The filter threshold was determined by the following process: first, 55 feature subsets of different sizes were generated by the DSDC-based feature selection method with different filter thresholds, then an MLP with two hidden layers was used to perform 10-fold cross-validation on each feature subset and the average accuracy was calculated. We chose the filter threshold corresponding to the highest average accuracy as the preset filter threshold (see Figure \ref{fig3}). Through DSDC-based feature selection, 3170 remarkable features were selected from the original 76636 features whose dimension was reduced by 95.86\%. Subsequent experimental results show that the feature selection not only improved the classifier's accuracy but also greatly reduced the training time of the deep learning model. In addition, the feature selection process of ABIDE I dataset with the input matrix size of 1035 x 76636 takes 28.12 seconds based on Intel Xeon Silver 4114 CPU, which reflects that the DSDC score is computationally efficient.

\subsection{Deep neural network classifier}
\label{s2.4}
This section introduces the architecture of our classifier and its training process.
\subsubsection{Simplified VAE and MLP\\}
\label{s2.4.1}
The training process of our classifier consists of simplified VAE unsupervised pretraining and MLP supervised fine-tuning.

VAE \cite{24, 25} is a generative model with an encoder and a decoder. Given an input vector x, the VAE's encoder outputs the latent space parameters ($\mu$ and $logvar$) by equations (\ref{eq:h}), (\ref{eq:mu}) and (\ref{eq:logvar}). Then the parameters are used to generate the latent variables z by equation (\ref{eq:z}). At last, VAE's decoder uses the latent variables to reconstruct the input vector.
\begin{equation}
\begin{aligned}[b]
h = W_{1}x + b_{1}
\label{eq:h}
\end{aligned}
\end{equation}
\begin{equation}
\begin{aligned}[b]
\mu = W_{2}h + b_{2}
\label{eq:mu}
\end{aligned}
\end{equation}
\begin{equation}
\begin{aligned}[b]
logvar = W_{3}h + b_{3}
\label{eq:logvar}
\end{aligned}
\end{equation}
\begin{equation}
\begin{aligned}[b]
z = \mu + \epsilon \times e^{0.5 \times logvar}
\label{eq:z}
\end{aligned}
\end{equation}
where $h$ is the output of the first hidden layer of VAE's encoder, $\mu$ and $logvar$ are parameters of the latent space, $\{W_{1},W_{2},W_{3},b_{1},b_{2},b_{3}\}$ are weights and biases of the encoder, $\epsilon$ is a random number sampled from $N(0,1)$ distribution.

We use the root-mean-square propagation (RMSProp) \cite{26} as the backpropagation method to optimize the loss function of VAE which is defined as follows,

\begin{equation}
\begin{aligned}[b]
Loss(W,b)= MAE + \beta\sum_{i=1}^n{KL[N(\mu_i , e^{logvar_i}) \parallel N(0,1)]}
\label{eq:loss}
\end{aligned}
\end{equation}
\begin{equation}
\begin{aligned}[b]
MAE=\frac{1}{n}\sum_{i=1}^{n}|f_{W,b}(x_i)-x_i|
\label{eq:mae}
\end{aligned}
\end{equation}
\begin{equation}
\begin{aligned}[b]
KL[N(\mu_i , e^{logvar_i}) \parallel N(0,1)]=-\frac{1}{2}(1+logvar_i-e^{logvar_i}-{\mu_i}^2)
\label{eq:kl}
\end{aligned}
\end{equation}

Where $Loss(W,b)$ represents the loss function of VAE. $MAE$ represents the mean absolute error \cite{27} between prediction and true label, $n$ is the number of samples, $x_i$ is the input of the ith sample, $f_{W,b}(x_i)$ is output of VAE's decoder. $KL[N(\mu_i , e^{logvar_i}) \parallel N(0,1)]$ represents the Kullback-Leibler divergence between $N(\mu_i , e^{logvar_i})$ and $N(0,1)$, $\mu_i$ and $logvar_i$ are parameters of latent space generated by encoder of VAE.

As seen from equations (\ref{eq:mu}) and (\ref{eq:logvar}), the $\mu$ and $logvar$ are generated by two branch networks with different parameters, respectively. However, our purpose is transfering the parameters of the pretrained VAE's encoder to MLP for fine-tuning, thus we simplified the structure of VAE's encoder by using a unified network to generate $\mu$ and $logvar$ simultaneously following equation (\ref{eq:s}) to ensure the encoder's structure is the same as the MLP's structure. 
\begin{equation}
\begin{aligned}[b]
\mu = logvar = W_{2}h + b_{2}
\label{eq:s}
\end{aligned}
\end{equation}

\begin{figure*}[h!]
\centering
\includegraphics[scale = 0.8]{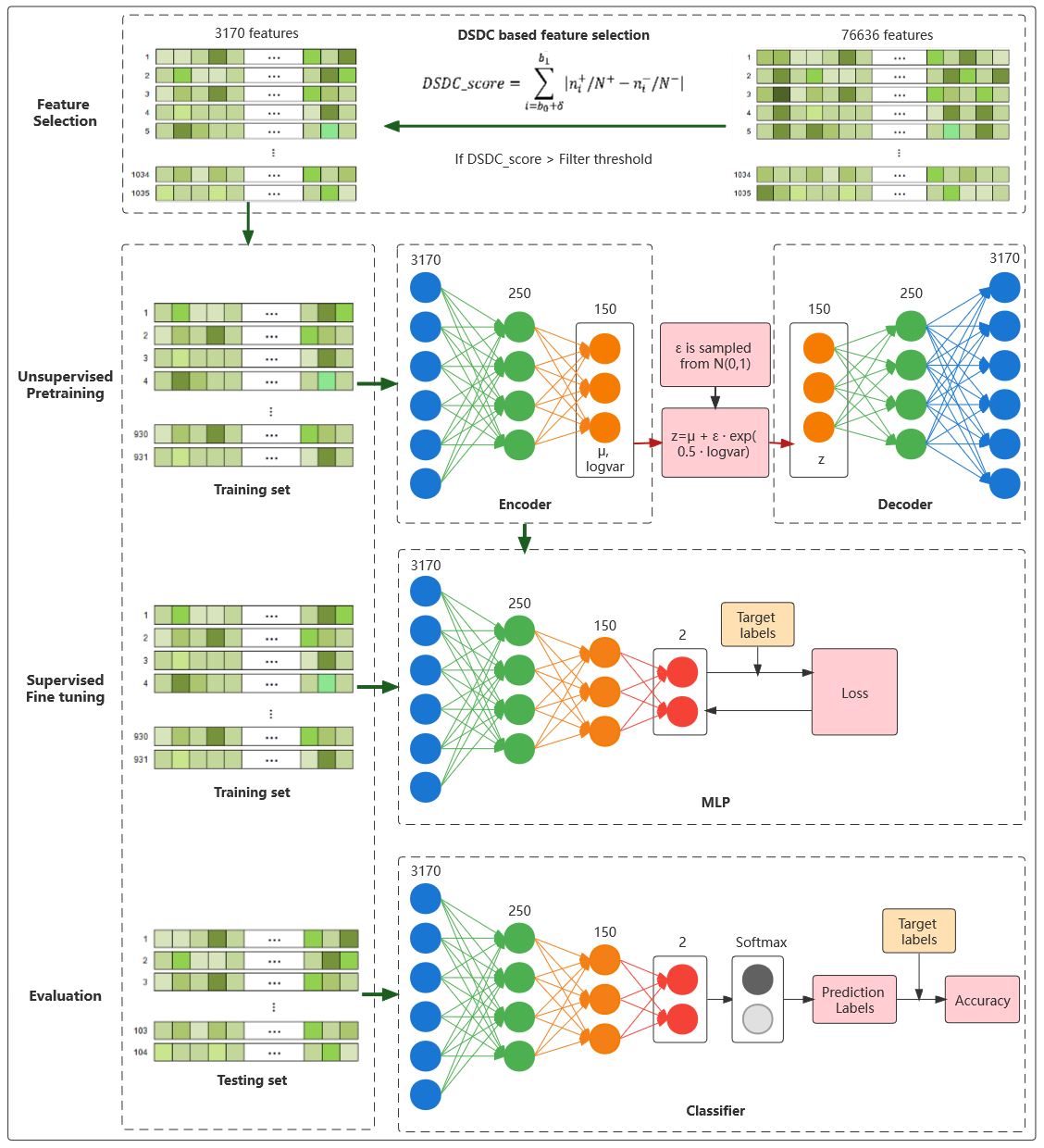} 
\caption{The entire process of feature selection, pretraining, fine-tuning and model evaluation}
\label{fig4}
\end{figure*}

After pretraining, we use the parameters of the VAE's encoder as the initialization parameters of the MLP and fine-tune the MLP's parameters through supervised training. Cross entropy is used as the loss function and RMSProp is adopted as the backpropagation method. A softmax layer is added after the MLP to calculate probabilities to determine the label of each subject. Afterwards, evaluation metrics such as accuracy, sensitivity, and specificity can be calculated by using the prediction label and true label. The model's architecture and the entire process of feature selection, pretraining, fine-tuning, and evaluation are shown in Figure \ref{fig4}. 

\subsubsection{Normalization and modified tanh activation function\\}
\label{s2.4.2}
To further improve the classification accuracy, a pipeline consisting of a normalization procedure and a modified tanh activation function was designed to replace the original tanh activation function. 

The normalization can be performed according to Equation (\ref{eq:norm}) through which the outputs of hidden layers are mapped to $[-1,1]$ in order to match the subsequent activation function. 
\begin{equation}
\begin{aligned}[b]
x_{norm}=\frac{2(x-x_{min})}{x_{max}-x_{min}}-1
\label{eq:norm}
\end{aligned}
\end{equation}
Where $x$ is the output of one hidden node and $x_{norm}$ is the normalized output; $x_{max}$ and $x_{min}$ is the maximum and minimum output of the hidden node. 

\begin{figure}[]
\centering
\renewcommand{\familydefault}{\sfdefault}\normalfont
\includegraphics[scale=0.45]{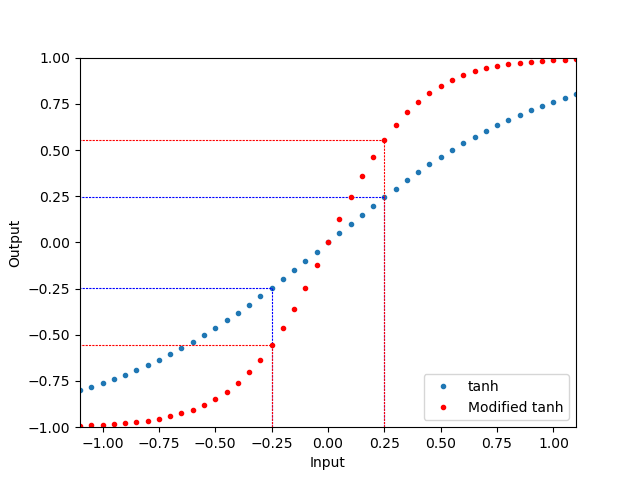}
\caption{A comparision of the tanh and the modified tanh}%
\label{fig5}
\end{figure}

After normalization, the modified tanh activation function in Equation (\ref{eq:modified_tanh}) is applied to the outputs of hidden layers. Compared with tanh, the modified tanh can make the features whose values close to zero more discriminating by mapping them to a larger interval (see Figure \ref{fig5})
\begin{equation}
\begin{aligned}[b]
y_{norm\_act}=\frac{e^{2.5 \cdot x_{norm}}-e^{-2.5 \cdot x_{norm}}}{e^{2.5 \cdot x_{norm}}+e^{-2.5 \cdot x_{norm}}}
\label{eq:modified_tanh}
\end{aligned}
\end{equation}
Where $y_{norm\_act}$ is the output of normalization-activation pipeline.

\subsubsection{Threshold moving\\}
\label{s2.4.3}
Threshold moving is an approach to alleviate the impact of class imbalance on classification performance by adjusting the classification threshold. When the data is balanced, the classification threshold of a softmax layer's output is usually set to 0.5. However, the number of HC samples is larger than that of ASD samples in ABIDE I, therefore, the threshold moving approach was adopted in our work and the decision is made according to the following rules. 
\begin{equation}
\begin{aligned}[b]
\frac{p_{ASD}}{p_{HC}} \cdot \frac{N_{HC}}{N_{ASD}} > 1 \Leftrightarrow \frac{p_{ASD}}{p_{HC}} >  \frac{N_{ASD}}{N_{HC}} \Rightarrow ASD
\label{eq:tmasd}
\end{aligned}
\end{equation}
\begin{equation}
\begin{aligned}[b]
\frac{p_{ASD}}{p_{HC}} \cdot \frac{N_{HC}}{N_{ASD}} < 1 \Leftrightarrow \frac{p_{ASD}}{p_{HC}} <  \frac{N_{ASD}}{N_{HC}} \Rightarrow HC
\label{eq:tmtc}
\end{aligned}
\end{equation}
Where $[p_{ASD},p_{HC}]$ are the outputs of the softmax layer, representing the probabilities of a sample being ASD and HC. $N_{HC}$ and $N_{ASD}$ are the number of HC and ASD samples in the training set, respectively.

As shown in equations (\ref{eq:tmasd}) and (\ref{eq:tmtc}), in the classification decision stage, we increase the weight of ASD which is equivalent to relaxing the criterion for determining ASD.

\subsection{Additional constraints during model training}
\label{s2.5}
This section will introduce two constraints that can help to train models with higher sensitivity and specificity, respectively. The constraints are used in the model training process to determine whether the optimized model parameters of a certain training epoch should be saved. The mechanism of the constraints is detailed in Algorithms \ref{algo:1}. 

Take constraint 1 for example, in the training process, constraint 1 helps to improve the difference between sensitivity and specificity while improving the accuracy. The sensitivity increases because it is positively correlated with the difference between sensitivity and specificity which can be proved as follows.\\\\
Because:
\begin{equation}
\begin{aligned}[b]
Accuracy=\frac{TP+TN}{N_{ASD}+N_{HC}}
\end{aligned}
\end{equation}
\begin{equation}
\begin{aligned}[b]
Sensistivity=\frac{TP}{N_{ASD}},\quad Specificity=\frac{TN}{N_{HC}}
\end{aligned}
\end{equation}
\begin{equation}
\begin{aligned}[b]
Sensitivity-Specificity=\frac{TP}{N_{ASD}}-\frac{TN}{N_{HC}}
\end{aligned}
\end{equation}
\begin{equation}
\begin{aligned}[b]
N_{ASD},N_{HC}=Constant
\end{aligned}
\end{equation}
Assume: 
\begin{equation}
\begin{aligned}[b]
Accuracy = Constant
\end{aligned}
\end{equation}
Then: 
\begin{equation}
\begin{aligned}[b]
TP+TN = Constant
\end{aligned}
\end{equation}
Obviously: 
\begin{equation}
\begin{aligned}[b]
\frac{TP}{N_{ASD}}-\frac{TN}{N_{HC}} \uparrow \Rightarrow TP \uparrow \Rightarrow Sensistivity \uparrow
\end{aligned}
\end{equation}
But if the difference between sensitivity and specificity keeps increasing and exceeds a certain value, the model's performance will deteriorate. Therefore an appropriate threshold is needed and the selection of the threshold will be discussed in detail in Section \ref{s3.3}.

\begin{algorithm}
\caption{}     
\small
\label{algo:1}       
\begin{algorithmic}[1] 
\Require \textit{1.initial model} \Comment Untrained Model with random initialization of parameters   

	\textit{2.training set}   \Comment The dataset used to train a model 

	\textit{3.validation set}  \Comment The dataset used to stop training automatically

	\textit{4.constraint\_type}  \Comment Parameter used to choose Constraint 1 or Constraint 2 

	\textit{5.threshold}   \Comment A preset threshold for constraints, the effect of which will be disscussed in Section \ref{s3.4}
\Ensure \textit{final model}  \Comment The trained model  
\State $max\_acc = 0$
\State \textit {$\delta$} $ = -1$    
\For {$training\_epoch = 1$ to $max\_training\_epoch$}
\State \textit{Training model on training set $\to$ current model}
\State \textit{Calculating accuracy, sensitivity and specificity of current model on validation set $\to$ $v\_acc, v\_sen, v\_spe$}
\If {$contraint\_type == 1$} 	\Comment Constraint 1
	\If {$v\_acc \ge max\_acc$ \textbf{and} $v\_sen - v\_spe \ge$ \textit{$\delta$}}
		\State $max\_acc = v\_acc$
		\If {\textit{$\delta$} $< threshold$}
			\State \textit{$\delta$} $= v\_sen - v\_spe$
		\Else
			\State \textit{$\delta$} $= threshold$
		\EndIf
		\State \textit{final model = current model}
	\EndIf
\ElsIf {$contraint\_type == 2$} \Comment Constraint 2
	\If {$v\_acc \ge max\_acc$ \textbf{and} $v\_spe - v\_sen \ge$ \textit{$\delta$}}
		\State $max\_acc = v\_acc$
		\If {\textit{$\delta$} $< threshold$}
			\State \textit{$\delta$} $= v\_spe - v\_sen$
		\Else
			\State \textit{$\delta$} $= threshold$
		\EndIf
\State \textit{final model = current model}
\EndIf
\Else
	\If {$v\_acc \ge max\_acc$ \textbf{and} $|v\_sen - v\_spe| \le$ \textit{$\delta$}}
		\State $max\_acc = v\_acc$
		\If {\textit{$\delta$} $> threshold$}
			\State \textit{$\delta$} $= |v\_sen - v\_spe|$
		\Else
			\State \textit{$\delta$} $= threshold$
		\EndIf
		\State \textit{final model = current model}
	\EndIf
\EndIf
\EndFor
\State\Return \textit{final model} 
\end{algorithmic} 
\end{algorithm} 


\section{RESULTS AND DISCUSSION}
\label{s3}
On ABIDE I dataset (505 ASD / 530 HC), most previous studies evaluated their models through a single 10-fold cross-validation, the evaluation results of which may be susceptible to the randomness of the dataset division. In our work, 10-fold cross-validation was repeated 10 times to evaluate our model more objectively. For each 10-fold cross-validation, we performed a stratified sampling to randomly divide the dataset into a training set, a validation set, and a testing set in an 8:1:1 ratio. The training set was used to train a model and the validation set was used to stop training automatically when the validation accuracy stopped increasing to avoid overfitting, and the testing set is used for evaluating the classification performance of the trained model. Accuracy, sensitivity, specificity, and training time are obtained by calculating the mean value of 10 times experiments for model evaluation. A grid search with a step size of 50 was performed to optimize the model's layer configuration from full[100]-full[100]-full[2] to full[1000]-full[900]-full[2], through which the layer configuration is determined as full[250]-full[150]-full[2], corresponding to the highest accuracy (78.12\%), where full[N] denotes a fully-connected layer with N outputs. 

In the following subsections, first, the DSDC-based feature selection method is compared with two other commonly used feature selection methods to highlight the advantage of the former. Next, the contribution of each procedure in our framework to classification accuracy is discussed. Then, we analyze the model performance by using constraints with different thresholds and provide a feasible threshold selection range. After that, our experimental results are compared with other state-of-the-art studies on the same dataset. Finally, the limitations of the current work are discussed. 

\subsection{Evaluation of feature selection methods}
\label{s3.1}

\begin{figure}[]
\centering
\includegraphics[scale=0.45]{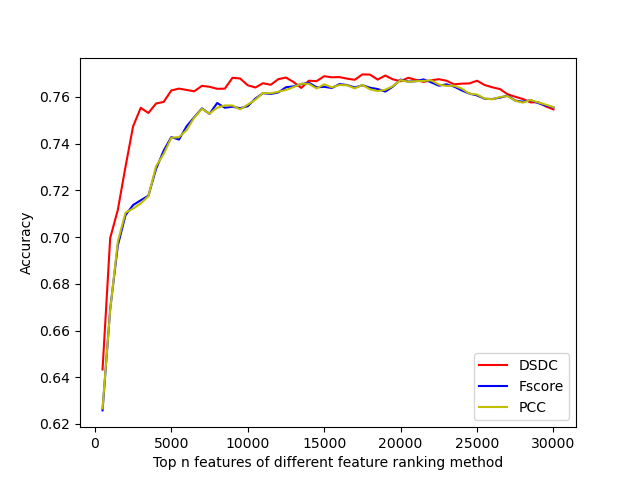}
\caption{A comparision of different feature selection methods based on the same SVM classifier}
\label{fig6}
\end{figure}

The advantage of the proposed DSDC-based feature selection method is highlighted by comparing it with two widely used feature selection methods based on F-score  \cite{Chen2006} and PCC. At first, all features are ranked by DSDC score, F-score, and the absolute value of PCC in descending order. Then top n (n varies from 500 to 30000) features of the different feature rankings were fed into the same SVM classifier. Finally, we compared the accuracy of the SVM classifier under different feature rankings (Figure  \ref{fig6}).

As in Figure \ref{fig6}, the red curve (representing DSDC) rises faster, indicating the top features of DSDC-based feature selection include more vital information than the top features of the other two methods (F-score and PCC). With the increase in the number of top features, more important features have been selected by all the three methods so that the accuracy difference between the three methods becomes smaller, while the DSDC-based feature selection method still showed better performance. This illustrates that DSDC can help to select more refined features and contribute to better classification performance. It is worth mentioning that we use an SVM classifier to compare different feature selection methods instead of a deep learning model because the model parameters will change according to different input dimensions. Changes in the parameters of a deep learning model commonly have an influence on the model performance, and it is difficult to tell whether changes in classification performance are mainly caused by the new inputs or the changes in model parameters.


\begin{figure}[]
\includegraphics[scale =0.6]{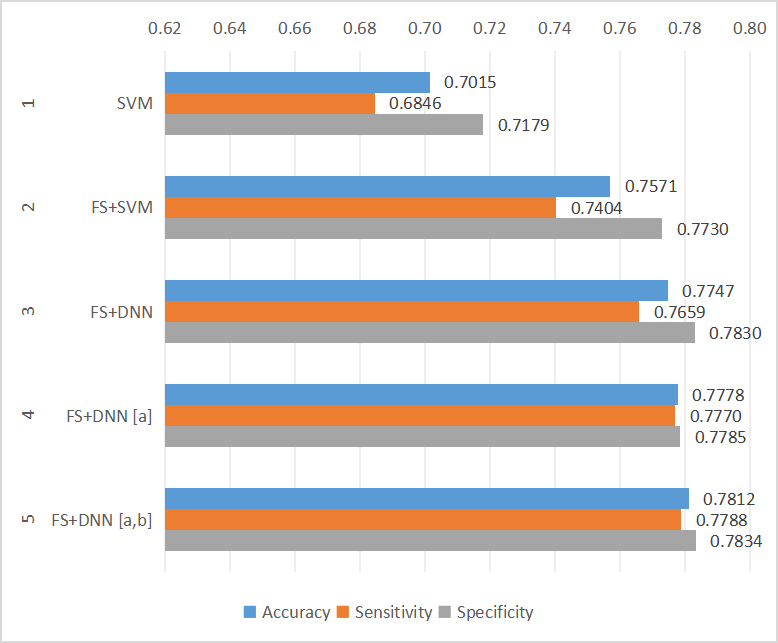}
\caption{Performance comparison of SVM, FS+SVM, FS+DNN, FS+DNN[a] and FS+DNN[a,b] models}
\begin{itemize}
    \scriptsize
    \item FS: The DSDC-based feature selection described in Section \ref{s2.3}
    \item DNN: MLP pretrained by simplified VAE described in Section \ref{s2.4.1}
    \item a: Using normalization and modified tanh activation function described in Section \ref{s2.4.2} instead of tanh in DNN
    \item b: Using threshold moving approach described in Section \ref{s2.4.3} in DNN
\end{itemize}
\label{fig7}
\end{figure}

\subsection{The performance analysis of our framework}
\label{s3.2}
The contribution to the classification accuracy of each procedure in our framework is shown in Figure \ref{fig7}. Besides, we performed two-sample t-tests and calculated p-values to investigate the significance of accuracy improvement (p-value $< 0.05$ indicates a significant improvement in accuracy). Due to the feature vectors without dimensionality reduction can bring a huge time cost to the training and hyperparameter tuning of a deep learning model, an SVM classifier is used instead to evaluate the performance improvement contributed by DSDC-based feature selection (see the 1st and 2nd rows). It can be observed that the DSDC-based feature selection resulted in a significant accuracy improvement of 5.56\% (p-value $< 0.01$). The 2nd and 3rd rows verify the advantages of deep learning over traditional machine learning by comparing the deep learning model pretrained by simplified VAE with SVM (p-value $< 0.01$). By replacing the original tanh activation function with the normalization-activation pipeline described in section \ref{s2.4.2} and using the threshold moving approach described in section \ref{s2.4.3}, a 0.65\% increase in accuracy is observed by comparing the 3rd and 5th rows (p-value $=0.02$), indicating that using these methods can significantly improve the accuracy of our model, while the individual effects of the normalization-activation pipeline (0.31\% accuracy increase showed in 3th and 4th rows) and threshold moving (0.34\% accuracy increase showed in 4th and 5th rows) are slight. In the loss function of the simplified VAE, we have also tried to use MSE loss instead of MAE loss, whereas the accuracy of the final classifier slightly decreased by 0.28\%. One possible reason is that the MAE loss is more robust to outliers than the MSE loss, and more than 89\% of the 3170 input features have outliers if values more than three times the standard deviation away from the mean value is regarded as outliers.

\begin{figure}[]
\centering
\includegraphics[scale =0.45]{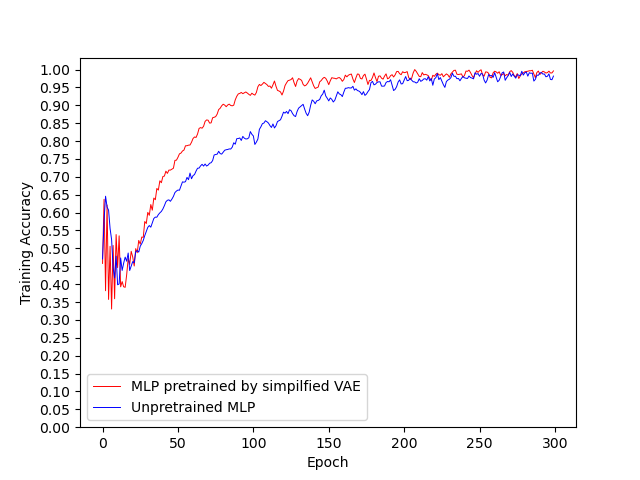}
\caption{A comparision of convergence speed between MLP pretrained by simplified VAE and unpretrained MLP}
\label{fig8}
\end{figure}

Figure \ref{fig8} was used to investigate the influence of the simplified VAE pretraining method on the convergence speed of the classifier by comparing the average training accuracy of the 10 times 10 folds' experiments after each training epoch. As shown in Figure \ref{fig8}, the red curve (representing the MLP pretrained by simplified VAE) is above the blue curve (representing the unpretrained MLP), which shows that the simplified VAE pretraining can speed up the convergence of the MLP by providing it with initialization parameters.

\begin{figure}[]
\centering
\includegraphics[scale =0.55]{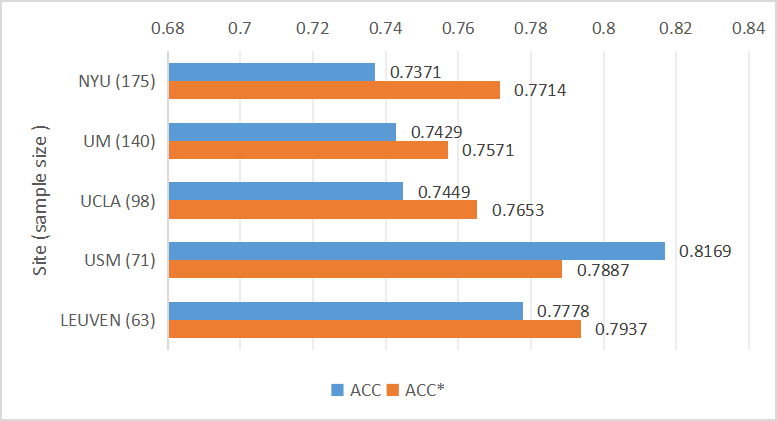}
\caption{Accuracy comparision of MLPs pretrained independently on the five largest sites of ABIDE I dataset and pretrained on the entire ABIDE I dataset}
\begin{itemize}
    \scriptsize
    \item ACC: The accuracy of MLP pretrained and fine-tuned on a single site
    \item ACC*: The accuracy of MLP pretrained on the entire ABIDE I dataset and fine-tuned on a single site
\end{itemize}
\label{fig9}
\end{figure}

Since different sites of ABIDE I adopt different scanning protocols, there is heterogeneity among data from different sites. To investigate whether the simplified VAE pretraining method can extract more useful information from a large heterogeneous dataset. We selected the five largest sites from the ABIDE I dataset and performed pretraining, fine-tuning, and inference independently on each of them (the corresponding accuracy is represented by blue bars in Figure \ref{fig9}). For comparison, we used the entire ABIDE I dataset (excluding the testing set) for pretraining and performed fine-tuning and inference independently on each of the 5 sites (the corresponding accuracy is represented by orange bars in Figure \ref{fig9}). As shown in Figure \ref{fig9}, for LEUVEN, UCLA, UM, and NYU sites, classifiers pretrained on the entire ABIDE I dataset achieved higher accuracy than those pretrained on a single site, which indicates that in most cases more useful patterns could be extracted from a larger heterogeneous dataset through the simplified VAE. However, the result of the USM site is an exception. One possible reason is the features' distribution the USM samples has a relatively bigger difference from that of other sites.

\subsection{The effects of constraints on model performance}
\label{s3.3}
\begin{figure}[]
\centering
\includegraphics[scale=0.48]{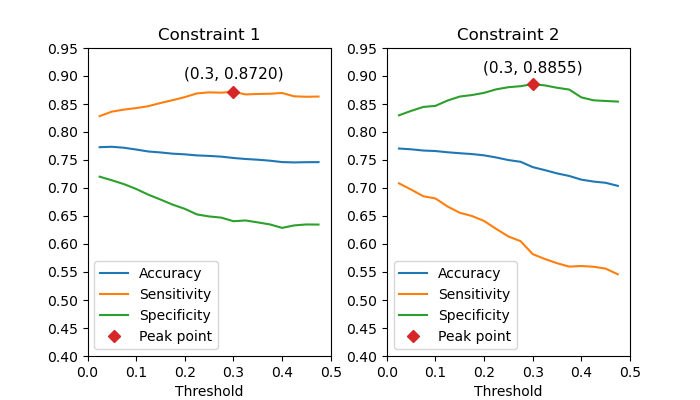}
\caption{Influence of different thresholds for constraints in Algorithm \ref{algo:1} on accuracy, sensitivity and specificity of classifier}
\label{fig10}
\end{figure}

\begin{figure*}[]
\centering
\includegraphics[scale=0.6]{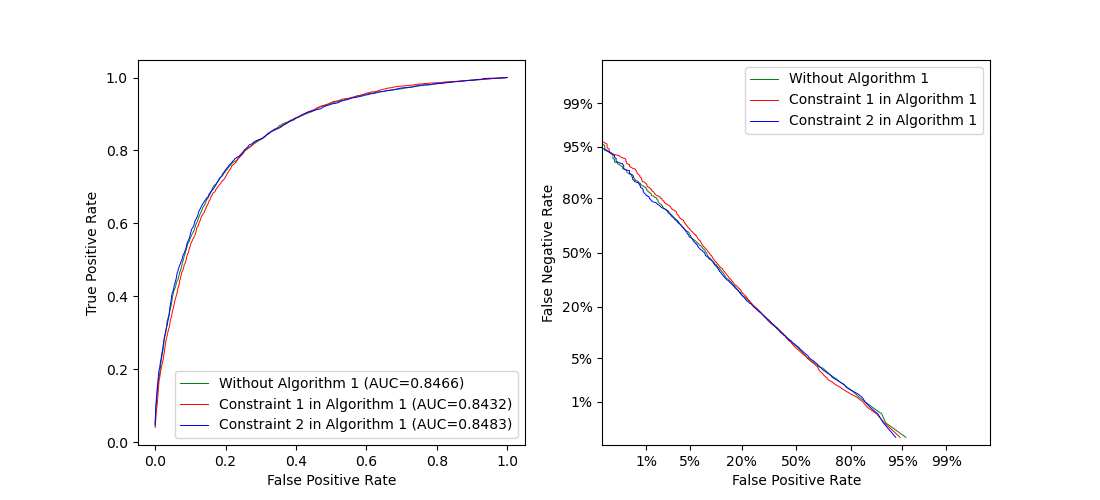}
\caption{ROCs (left) and DET curves (right) of classifiers trained with and without constraints in Algorithm \ref{algo:1}}
\begin{itemize}
    \scriptsize
    \item The threshold of the constraints is set to 0.3
\end{itemize}
\label{fig11}
\end{figure*}

Section \ref{s2.5} has described the mechanism of the two constraints used to train models with higher sensitivity and specificity and explained that an appropriate threshold is necessary to ensure a good model performance. This section will analyze the influence of constraints' thresholds on model performance. Our experimental results demonstrate that the constraints can significantly improve the model's sensitivity or specificity at a cost of small accuracy reduction while ensuring the overall performance. Figure \ref{fig10} summarizes the trade-off among accuracy, sensitivity, and specificity under different thresholds in Algorithm \ref{algo:1}. Take constraint 1 as an example, in the range of 0 to 0.3, selecting a higher threshold will improve the sensitivity, but at a cost, the accuracy and specificity will decrease; When the threshold exceeds 0.3, the performance of the model deteriorates as the threshold increases. Constraint 2 follows a similar pattern. Therefore, the value range for the threshold is 0 to 0.3. In our work, we set the threshold to be 0.3 and trained two models with 87.20\% sensitivity and 88.55\% specificity by using constraint 1 and constraint 2 in Algorithm \ref{algo:1}, respectively. Compared with the model trained without using the two constraints (Model 1 in Table \ref{Table 3}), Model 2 in Table \ref{Table 3} improved sensitivity by 9.32\% at the cost of an accuracy reduction of 2.76\%  and Model 3 in Table \ref{Table 3} improved specificity by 10.21\% at the cost of an accuracy reduction of 4.38\%. For disease diagnosis, a model with high sensitivity and specificity can reduce missed diagnosis and misdiagnosis rates which reflects our model could potentially adapt to the different cases for clinical application. For example, a model with higher sensitivity is very useful for the cases like COVID-19 screening since missing the diagnosis of a virulent communicable illness has far-reaching public health implications and may accelerate the pandemic. Models with higher sensitivity and specificity can also be used to perform a double check on the doctor's diagnosis results. For subjects whose doctor's diagnosis results are inconsistent with the model's prediction results, further analysis or detection can be performed to eliminate the potential diagnostic errors. In order to evaluate the influence of Algorithm \ref{algo:1} on the model's overall performance, Figure \ref{fig11} compares the models' average AUCs, ROCs, and DET curves before and after using Algorithm \ref{algo:1}. The result shows that the three models have similar ROCs and AUCs which demonstrates that the proposed constraints can ensure the model's overall performance while improving sensitivity or specificity. In Figure \ref{fig11}, the DET curves of the classifier trained without using Algorithm \ref{algo:1} and classifiers trained with constraint 1 and constraint 2 in Algorithm \ref{algo:1} are represented by green, red, and blue curves, respectively. It can be observed that the red curve is generally below the green curve when the false-negative rate is lower than 20\% and the blue curve is generally below the green curve when the false-positive rate is lower than 20\%. This indicates that when the false-negative rate and the false-positive rate are relatively low, constraint 1 and constraint 2 can contribute to better model performance, respectively.

\begin{sidewaystable*}

\renewcommand\arraystretch{1.5}  
\scriptsize
\caption{Performance comparison between our work and other state-of-the-art studies }
\vspace{8pt}
\centering
\begin{tabular}{|l|c|c|c|c|c|c|c|c|c|c|c|c|c|}
\hline

\multirow{2}{*}{ }&\multirow{2}{*}{\textbf{Dataset}}&\multirow{2}{*}{\textbf{Model}}&\multirow{2}{*}{\textbf{Validation}}&\multirow{2}{*}{\textbf{ACC}}&\multirow{2}{*}{\textbf{SEN}}&\multirow{2}{*}{\textbf{SPE}}&\multicolumn{3}{|c|}{\tabincell{c}{\textbf{Best cases in }\\\textbf{10 times 10-fold CV}}}&\multicolumn{3}{|c|}{\tabincell{c}{\textbf{Worst cases in }\\\textbf{10 times 10-fold CV}}}&\multirow{2}{*}{\tabincell{c}{\textbf{Training time}\\ \textbf{of 10-fold CV}}}\\\cline{8-13}

& & & & & & & \textbf{ACC} & \textbf{SEN} & \textbf{SPE}  & \textbf{ACC} & \textbf{SEN} & \textbf{SPE} & \\
 
\hline
Model 1 (Our proposed)& \tabincell{c}{ABIDE I  \\ 505 ASD \& \\ 530 HC}	 &	\tabincell{c}{Simplified VAE + MLP} &	\tabincell{c}{Mean value of \\10 times 10-fold CV} &  \textit{\textbf {0.7812}} &	0.7788 &	0.7834 &	\textit{\textbf {0.7894}} &	0.804 &	0.8019 &	0.77 &	0.7564 &	0.7528 &	\tabincell{c}{85s \\ NVIDIA Tesla P100}\\

\hline
Model 2 (Our proposed)& \tabincell{c}{ABIDE I  \\ 505 ASD \& \\530 HC}	 &	\tabincell{c}{Simplified VAE + MLP \\ussing constraint 1 in Algorithm \ref{algo:1}} &	\tabincell{c}{Mean value of \\10 times 10-fold CV} & 0.7536 &	\textit{\textbf {0.8720}} &	0.6408 &	0.7739&	\textit{\textbf {0.895}}&	0.6943&	0.7391&	0.8535&	0.6038&	\tabincell{c}{85s \\ NVIDIA Tesla P100}\\

\hline
Model 3 (Our proposed)& \tabincell{c}{ABIDE I  \\ 505 ASD \& \\ 530 HC}	 &	\tabincell{c}{Simplified VAE + MLP \\using constraint 2 in Algorithm \ref{algo:1}} &	\tabincell{c}{Mean value of \\10 times 10-fold CV} & 0.7374 &	0.5820 &	\textit{\textbf {0.8855}} &	0.7536&	0.6198&	\textit{\textbf {0.9113}}&	0.7043&	0.5347&	0.866&	\tabincell{c}{85s \\ NVIDIA Tesla P100}\\

\hline
Huang et al. \cite{13} & \tabincell{c}{ABIDE I  \\ 505 ASD \& \\530 HC} &	\tabincell{c}{DBN with \\DE-based optimizer}&	10-fold CV  &	\tabincell{c}{0.7640 \\$\pm 0.022$} & $-$ &	$-$ &	$-$ &	$-$ &	$-$ &	$-$ &	$-$ &	$-$ &	$-$\\

\hline
Shrivastava et al. \cite{28} & \tabincell{c}{ABIDE I  \\ 505 ASD \& \\530 HC}	 & CNN &	10-fold CV &0.7602 &	0.7004 &	0.8169 &	$-$ &	$-$ &	$-$ &	$-$ &	$-$ &	$-$ &	$-$\\

\hline
Jiao et al. \cite{29} & \tabincell{c}{ABIDE I  \\ 505 ASD \& \\530 HC}	&	CapsNet &	10-fold CV & 0.7100& 	0.7300& 	0.6600& 	$-$ &	$-$ &	$-$ &	$-$ &	$-$ &	$-$ &	$-$\\

\hline
Eslami et al. \cite{12} & \tabincell{c}{ABIDE I  \\ 505 ASD \& \\530 HC}	 &	ASD-DiagNet & 10-fold CV &0.7030& 	0.6830& 	0.7220& 	$-$ &	$-$ &	$-$ &	$-$ &	$-$ &	$-$ &	\tabincell{c}{41 min\\NVIDIA Tesla K40c}\\

\hline
Sherkatghanad et al. \cite{30} & \tabincell{c}{ABIDE I  \\ 505 ASD \& \\530 HC}	&	CNN &	10-fold CV &	0.7022 &	0.7746 &	0.6182& 	$-$ &	$-$ &	$-$ &	$-$ &	$-$ &	$-$ &	\tabincell{c}{12h and 30min \\ NVIDIA Tesla K80}\\

\hline
Heinsfeld et al. \cite{11} & \tabincell{c}{ABIDE I  \\ 505 ASD \& \\530 HC}	&	SAE & 10-fold CV & 0.7000& 	0.7400& 	0.6300& 	$-$ &	$-$ &	$-$ &	$-$ &	$-$ &	$-$ &	\tabincell{c}{32h 52 m 36 s \\  NVIDIA Tesla K40}\\

\hline
\end{tabular}
\label{Table 3}
\end{sidewaystable*}

\subsection{Comparision with other state-of-the-art methods on the same dataset}
\label{s3.4}	
In this section, we make a comparison of our proposed method with several previous methods based on the same dataset in Table \ref{Table 3}. The state-of-the-art methods used for comparison include denoising autoencoder, convolutional neural network (CNN), DBN, CapsNet, and ASD-Diagnet, all of which have achieved over 70\% prediction accuracy. In our work, we trained three models based on simplified VAE and MLP, Model 2 and 3 were trained by using the constraint 1 and constraint 2 described in algorithm \ref{algo:1}, respectively, while Model 1 is trained without using the two constraints. The accuracy (78.12\%), sensitivity (87.20\%), and specificity (88.55\%) of Model 1, Model 2, and Model 3 exceed the corresponding results in previous studies on the same dataset, respectively. This highlights the outstanding classification performance and flexibility of our framework. Our method also runs efficiently due to less number of selected features. The last column of Table \ref{Table 3} lists the training time and the corresponding GPU used for training. The training time of our model is 85s for 10-fold cross-validation (8.5 seconds for training a single model), which has an advantage over other studies after taking into account the GPUs' performance differences. A model with a fast training speed can be retrained in a short time as new subjects arrive, which means that the model can potentially achieve real-time optimization and improvement in practical use. 

\subsection{Limitations}
\label{s3.5}	
In the present study, the experimental results are based on the ABIDE I dataset. More ASD datasets or other neurological diseases are expected to be used to evaluate the classification framework in the future. In addition, two other modalities of MRI (structural MRI and diffusion tensor imaging) have not been used in our work. Since different MRI modalities contain complementary information for ASD identification, fusing multiple modalities for ASD classification may work better than just using rs-fMRI. The proposed deep learning model can potentially be used to discover vital functional connectivities by studying its explainability, which can help to understand the ASD mechanism. However, this has not been included in the current work. 

\section{Conclusion}
\label{s4}
In this study, we proposed a novel filter feature selection method (DSDC-based feature selection method) to select remarkable FCs and designed a deep learning model with two procedures -- simplified VAE pretraining and MLP fine-tuning. In our model, we designed a pipeline consisting of normalization and a modified tanh activation function to replace the original tanh function and adopted the threshold moving approach, which can further improve the classification accuracy. In addition, we proposed two constraints that can help to train models with higher sensitivity or specificity. The outstanding classification performance on the heterogeneous dataset and adjustable sensitivity and specificity suggest that our method goes one step further based on state-of-the-art methods and has the potential to be a viable auxiliary method for ASD early detection. 

Future work will focus on fusing multimodal MRI data to utilize the complementary information for ASD identification, which is expected to further improve the classification performance, and experiments will be performed on more datasets to verify the robustness of the model. In addition, further work will be carried out to study the explainability of the proposed model, which can potentially be used to discover vital functional connectivities and help to understand the ASD mechanism.

\section*{Acknowledgments}
This work was partly supported by the Shenzhen KQTD Project No. KQTD20200820113106007, National Key Research and Development Program of China under Grant No. 2018YFB0204403, Strategic Priority CAS Project XDB38050100, National Science Foundation of China under grant no. U1813203, the Shenzhen Basic Research Fund under grant no. RCYX2020071411473419 and JSGG20201102163800001, CAS Key Lab under grant no. 2011DP173015. 

%
%
%
\bibliographystyle{unsrt} 
\bibliography{reference}


\end{document}